\documentclass[twocolumn,showpacs,superscriptaddress,preprintnumbers,amsmath,amssymb]{revtex4}

\usepackage{graphicx}
\usepackage{dcolumn}
\usepackage{bm}

\usepackage{color}
\usepackage{CJK}

\begin{document}

\preprint{1}

\title{Production of Kaon and $\Lambda$ in nucleus-nucleus collisions at  ultra-relativistic energy
from a blast wave model }

\author{S. Zhang}
\affiliation{Shanghai Institute of Applied Physics, Chinese Academy of Sciences, Shanghai 201800, China}
\author{Y. G. Ma\footnote{Author to whom all correspondence should be addressed: ygma@sinap.ac.cn}}
\affiliation{Shanghai Institute of Applied Physics, Chinese Academy of Sciences, Shanghai 201800, China}
\affiliation{Shanghai Tech University, Shanghai 200031, China}
\author{J. H. Chen}
\affiliation{Shanghai Institute of Applied Physics, Chinese Academy of Sciences, Shanghai 201800, China}
\author{C. Zhong}
\affiliation{Shanghai Institute of Applied Physics, Chinese Academy of Sciences, Shanghai 201800, China}

\date{\today}

\begin{abstract}
The particle production of Kaon and $\Lambda$ are studied in nucleus-nucleus collisions at relativistic energy based on a chemical equilibrium blast-wave model. The transverse momentum spectra of Kaon and $\Lambda$ at the kinetic freeze-out stage from our model are in good agreement  with the experimental results. The kinetic freeze-out parameters of temperature ($T_{kin}$) and radial flow parameter $\rho_{0}$ are presented for the FOPI, RHIC and LHC energies. And the resonance decay effect is also discussed. The systematic study for beam energy dependence of the strangeness particle production will help us to better understand the properties of the matter created in heavy-ion collisions at the kinetic freeze-out stage.
\end{abstract}

\maketitle

\section{Introduction}

Study for Quark-Gluon Plasma (QGP)~\cite{QCD-QGP} by Ultra-relativistic heavy-ion collisions~\cite{RHICWithePaper-1,RHICWithePaper-2,RHICWithePaper-3,RHICWithePaper-4} provides an opportunity to understand the production mechanism of strangeness hadrons and the phase space evolution in the system created, especially the  energy dependence is thought as one of important observables to study various aspects of the QCD
phase diagram \cite{Tian,LiuFM,Ko} in the beam energy scan  program at RHIC \cite{Mohanty}.
The s-quarks in strangeness hadrons can be a good probe to investigate the properties of the high dense and temperature matter because there should be no net strangeness content in the colliding system \cite{Phi,MaYG,Zhang2010,ChenJH2008}. The s-quarks are all created in the reaction. Kaon is produced associated with $\Lambda$-hyperon at FOPI energy, and at higher energies, such as in the RHIC or LHC region,  it can be produced by another mechanism, the so-called pair production~\cite{RHICWithePaper-1,RHICWithePaper-2,RHICWithePaper-3,RHICWithePaper-4}. The transverse momentum $p_{T}$ distribution encodes rich physics information. It has been measured in p + p, p + nuclues or nucleus + nucleus collisions at different energy range by the FOPI~\cite{FOPI-KL} and KaoS~\cite{KaoS-KL} collaboration at $~\sqrt{s_{NN}}=2.6 $ GeV, by the RHIC-STAR collaboration at $\sqrt{s_{NN}} = $ 62 GeV~\cite{RHIC-STAR-SYS-KL62,RHIC-STAR-KL62} and 200 GeV~\cite{RHIC-STAR-KL200-1,RHIC-STAR-KL200-2} and by the LHC-ALICE collaboration~\cite{LHC-ALICE-KL-1,LHC-ALICE-KL-2,LHC2-1,LHC2-2} at $\sqrt{s_{NN}} = $ 2.76 and 7 TeV etc. The resonance decay effect is also discussed in the above papers~\cite{RHIC-STAR-SYS-KL62, SONG-LHC-PRC}. 
Through fitting the measured transverse momentum $p_{T}$ distribution by a blast wave model~\cite{BLWave}, one can investigate the properties of the collision system at kinetic freeze-out stage. The temperature $T_{kin}$ parameter and the radial flow at kinetic freeze-out stage can provide some information about the evolution of the expanding phase space. One can also study if the system has reached kinetic equilibrium stage when the hadron interactions completed. The resonances decay will change the production rate and the $p_{T}$ distribution of Kaon and $\Lambda$. However, it is difficult to estimate the effect quantitatively in experiment. The blast wave model has been applied in experimental analysis~\cite{ALICE-chPCDepen,RHIC-STAR-SYS-KL62} to study the kinetic freeze-out properties. Retie\`ere and Lisa~\cite{BLWave-Fabrice} have explored in detail an analytic parameterization of the freeze-out configuration and investigated the spectra, the collective flow, and the HBT correlation of the hadrons produced in head-on nuclear collisions at top RHIC energy. In addition, the DRAGON~\cite{DRAGON} and the THERMINATOR2~\cite{THERMINATOR-1,THERMINATOR-2} models have been developed to study the phase-space distribution of produced hadrons at freeze-out stage. In this paper, we carry out a detailed study on the beam energy dependence of Kaon and $\Lambda$ production based on a blast-wave model. The transverse momentum distributions of Kaon and $\Lambda$ are presented. The chemical and kinetic freeze-out parameters are consistent with the ones extracted from experimental data. The resonance decay effect is also investigated and an approximate estimation of the percent of Kaon and $\Lambda$ from resonance decay to total Kaon and $\Lambda$ is achieved.

\section{Blast-wave model with thermal equilibrium mechanism}

As discussed above, within the framework of the
blast-wave model, the fireball created in high-energy heavy-ion
collisions is assumed to be in local thermal equilibrium and expands
at a four-component velocity $u_{\mu}$. The phase-space distribution
of hadrons emitted from the expanding fireball can be expressed as a
Wigner function~\cite{BLWave-Fabrice,DRAGON,THERMINATOR-1,THERMINATOR-2}
\begin{widetext}
\begin{eqnarray}
S(x,p)d^{4}x = \frac{2s+1}{(2\pi)^{3}}m_{t}\text{cosh}(y-\eta)\text{exp}\left(-\frac{p^{\mu}u_{\mu}}{T_{kin}}\right)\Theta(1-\tilde{r}(r,\phi))H(\eta)\delta(\tau-\tau_{0})d\tau\tau d\eta rdrd\phi,
\label{eq:BLWWigner}
\end{eqnarray}
\end{widetext}
where $s$, $y$, and $m_{t}$ are  the spin, rapidity, and
transverse mass of the hadron, respectively, and $p_{\mu}$ is the four-component
momentum. Equation~(\ref{eq:BLWWigner}) is formulated in a Lorentz
covariant way, $r$ and $\phi$ are the polar coordinates, and $\eta$
and $\tau$ are the pseudorapidity and the proper time, respectively.
$\tilde{r}$ is defined as
\begin{equation}
\tilde{r} =
\sqrt{\frac{(x^{1})^{2}}{R^{2}}+\frac{(x^{2})^{2}}{R^{2}}},
\label{eq:tilde_r}
\end{equation}
with ($x^{1}, x^{2}$) standing for the coordinates in the transverse
plane and $R$ being the average transverse radius. The kinetic
freeze-out temperature $T_{kin}$ and the radial flow parameter
$\rho_{0}$ are important in determining the transverse momentum
spectrum. And the latter will affect the four-component velocity
field. Since we are currently interested in the $p_T$ spectrum at
mid-rapidity, the pseudorapidity distribution $H(\eta)$ is not
important. The $p_{T}$ spectrum can then be written as
\begin{equation}
\frac{dN}{2\pi p_{T}dp_{T}} = \int S(x,p)d^{4}x.
\label{eq:pTSpectra}
\end{equation}
A thermal equilibrium model is employed to calculated the chemical components of hadrons. 
The particle density of species
\textit{i} can be expressed as~\cite{Thermal-EQ-1,Thermal-EQ-2,Thermal-EQ-3,Thermal-EQ-4, Thermal-PMB-1,Thermal-PMB-2,DRAGON}
\begin{widetext}
\begin{eqnarray}
n_{i}(T_{ch},\mu_{B},\mu_{S})
&=& g_{i}\int \frac{d^{3}p}{(2\pi)^{3}}\left[\text{exp}\left(\frac{\sqrt{p^{2}+m_{i}^{2}}-\left(\mu_{B}B_{i}+\mu_{S}S_{i}\right)}{T_{ch}}\right)\mp1\right]^{-1}\nonumber\\
&=&I\left(g_i,\frac{m_i}{T_{ch}}\right)\sum_{n=1}(\pm 1)^{n+1}\exp{\left(n\frac{\left(\mu_{B}B_{i}+\mu_{S}S_{i}\right)}{T_{ch}}\right)},\nonumber\\
I\left(g_i,\frac{m_i}{T_{ch}}\right)&=&g_{i}\int\frac{d^{3}p}{(2\pi)^{3}}\left[\sum_{n=1}(\pm
1)^{n+1}\exp{\left(-n\frac{\sqrt{p^{2}+m_{i}^{2}}}{T_{ch}}\right)}\right],
\label{eq:ni}
\end{eqnarray}
\end{widetext}
with the upper (lower) sign for bosons (fermions) and $g_{i}$ being the degeneracy factor. Assuming that the chemical equilibrium condition is satisfied, Eq.~(\ref{eq:ni}) essentially determines the fraction of particle species \textit{i}. The particle yield can be described by adjusting parameters such as the chemical freeze-out temperature $T_{ch}$, the baryon chemical potential $\mu_{B}$, the strangeness chemical potential $\mu_{S}$, and the system volume $V$. And then the fraction of particle species $i$ and its phase-space distribution can be calculated from Eqs.~(\ref{eq:ni}) and (\ref{eq:pTSpectra}).

\section{Results and discussion}

\begin{figure}
\includegraphics[width=8cm]{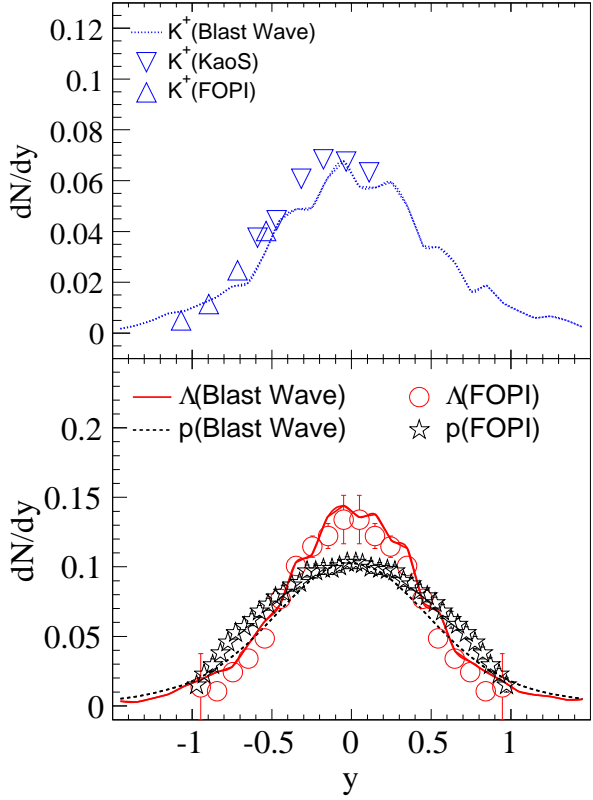}
\caption{\label{fig:rapidity} (Color online) Rapidity distribution of $K^{+}$, $\Lambda$ and proton (normalized to $\Lambda$ yield) in Ni + Ni central collisions at $\sqrt{s_{NN}}$ = 2.6 GeV. Lines represent the model calculations with $T_{kin}=30$ MeV, $\rho_{0}$=0.6, $T_{ch}$=49 MeV and $\mu_{B}$ = 815 MeV; Symbols are data from the FOPI~\cite{FOPI-KL} and KaoS~\cite{KaoS-KL} experiments.}
\end{figure}

The rapidity distribution of $K^{+}$, proton and $\Lambda$ are presented in figure~\ref{fig:rapidity}. The hadron rapidity distribution is related to the density at reaction scope. The calculated results can describe the data from the FOPI~\cite{FOPI-KL} and KaoS collaborations~\cite{KaoS-KL} for $K^{+}$, proton and $\Lambda$ in Ni+Ni collisions at $\sqrt{s_{NN}} \simeq$ 2.6 GeV. The $\Lambda$ is mainly produced at midrapidity (-0.5$<y<$0.5) and different from proton distribution at forward/backward rapidity. This results from the pronounced longitudinal expansion of proton and indicates large degree of transparency in Ni + Ni system at the FOPI energies ~\cite{FOPI-KL}. A similar calculation for $dN/dy$ from other groups can be found in Ref.~\cite{dNdy-NPA772-167} and our previous results for $\Lambda$ were calculated by a transport model~\cite{LNi-CPC-35-741}. 

The collective properties of the hot and dense matter created in
ultra-relativistic heavy-ion collisions at freeze-out stage can be
studied through transverse momentum ($p_{T}$) distributions of
identified particles. The radial flow
\begin{equation}
\label{eq-betaT}
\langle \beta \rangle = \int
\text{arctanh}\left(\rho_{0}\frac{r}{R}\right)rdr/\int rdr
\end{equation}
is related to the maximum flow rapidity
\begin{equation}
\rho = \tilde{r}\left[\rho_{0}+\rho_{a}\text{cos}(2\phi)\right].
\end{equation}
The results are independent of $\rho_{a}$ since the
$\phi$ dependence is averaged out after integration. 

\begin{figure}
\includegraphics[width=8cm]{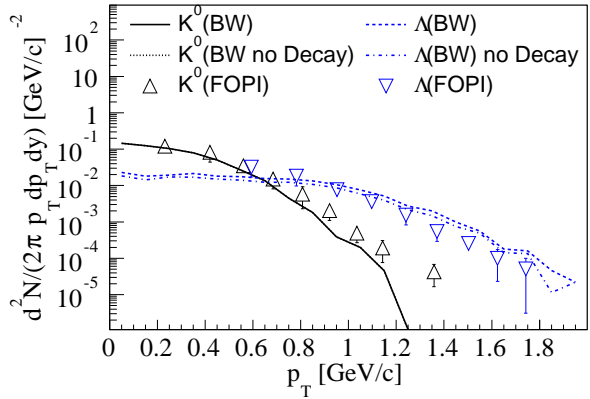}
\caption{\label{fig:pT-FOPI} (Color online) The transverse momentum $p_{T}$ distribution of $K^{0}$ and $\Lambda$ at mid-rapidity in Ni + Ni central collisions at $\sqrt{s_{NN}} \simeq$ 2.6 GeV. Lines represent the model calculations with $T_{kin}=30$ MeV, $\rho_{0}$=0.6, $T_{ch}$=49 MeV and $\mu_{B}$ = 815 MeV;
Symbols are data from the FOPI collaboration~\cite{FOPI-KL}.}
\end{figure}

\begin{figure}
\includegraphics[width=8cm]{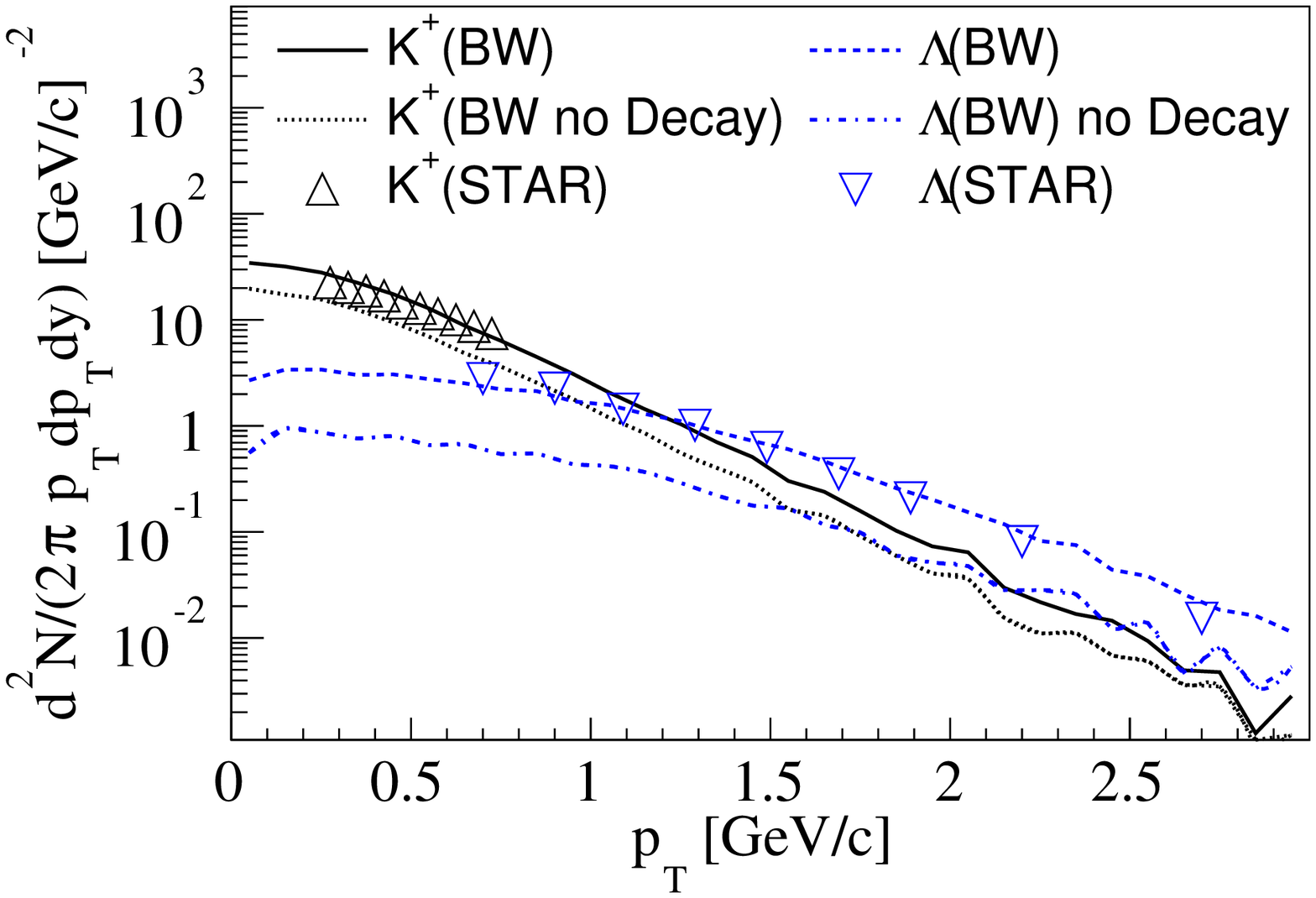}
\caption{\label{fig:pT-RHIC62} (Color online) The transverse momentum $p_{T}$ distribution of $K^{+}$ and $\Lambda$ at mid-rapidity in Au + Au central collisions at $\sqrt{s_{NN}}$ = 62.4 GeV. Lines represent the model calculations with $T_{kin}=105$ MeV, $\rho_{0}$=0.7, $T_{ch}$=156 MeV and $\mu_{B}$ = 70 MeV;
Symbols are data from the  RHIC-STAR collaboration~\cite{RHIC-STAR-SYS-KL62,RHIC-STAR-KL62}.}
\end{figure}

\begin{figure}
\includegraphics[width=8cm]{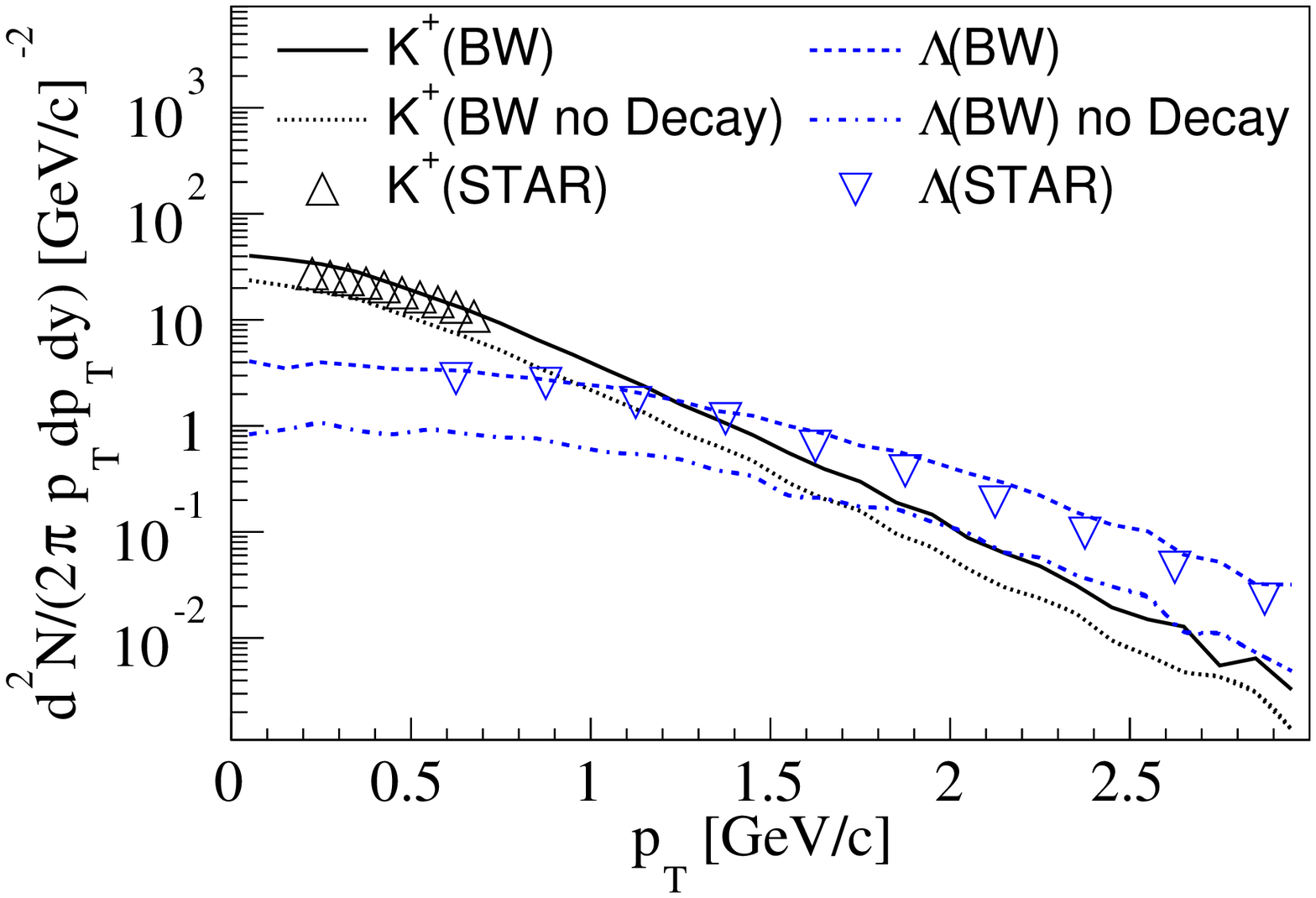}
\caption{\label{fig:pT-RHIC200} (Color online) The transverse momentum $p_{T}$ distribution of $K^{+}$ and $\Lambda$ at mid-rapidity in Au + Au central collisions at $\sqrt{s_{NN}}$ = 200 GeV. Lines represent the model calculations with $T_{kin}=91$ MeV, $\rho_{0}$=0.8, $T_{ch}$=156 MeV and $\mu_{B}$ = 20 MeV;
Symbols are data from the RHIC-STAR collaboration~\cite{RHIC-STAR-KL200-1,RHIC-STAR-KL200-2}.}
\end{figure}

\begin{figure}
\includegraphics[width=8cm]{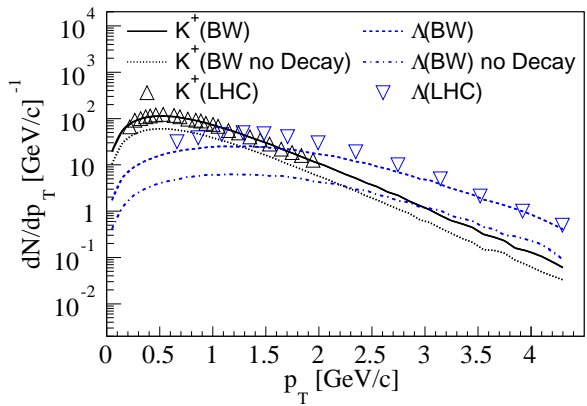}
\caption{\label{fig:pT-LHC} (Color online) The transverse momentum $p_{T}$ distribution of $K^{+}$ and $\Lambda$ at mid-rapidity in Pb + Pb central collisions at $\sqrt{s_{NN}}$ = 2.76 TeV. Lines represent the model calculations with $T_{kin}=99$ MeV, $\rho_{0}$=1, $T_{ch}$=160 MeV and $\mu_{B}$ = 10 MeV;
Experimental data are taken from Refs.~\cite{LHC-ALICE-KL-1,LHC-ALICE-KL-2}.}
\end{figure}

\begin{figure}
\includegraphics[width=8cm]{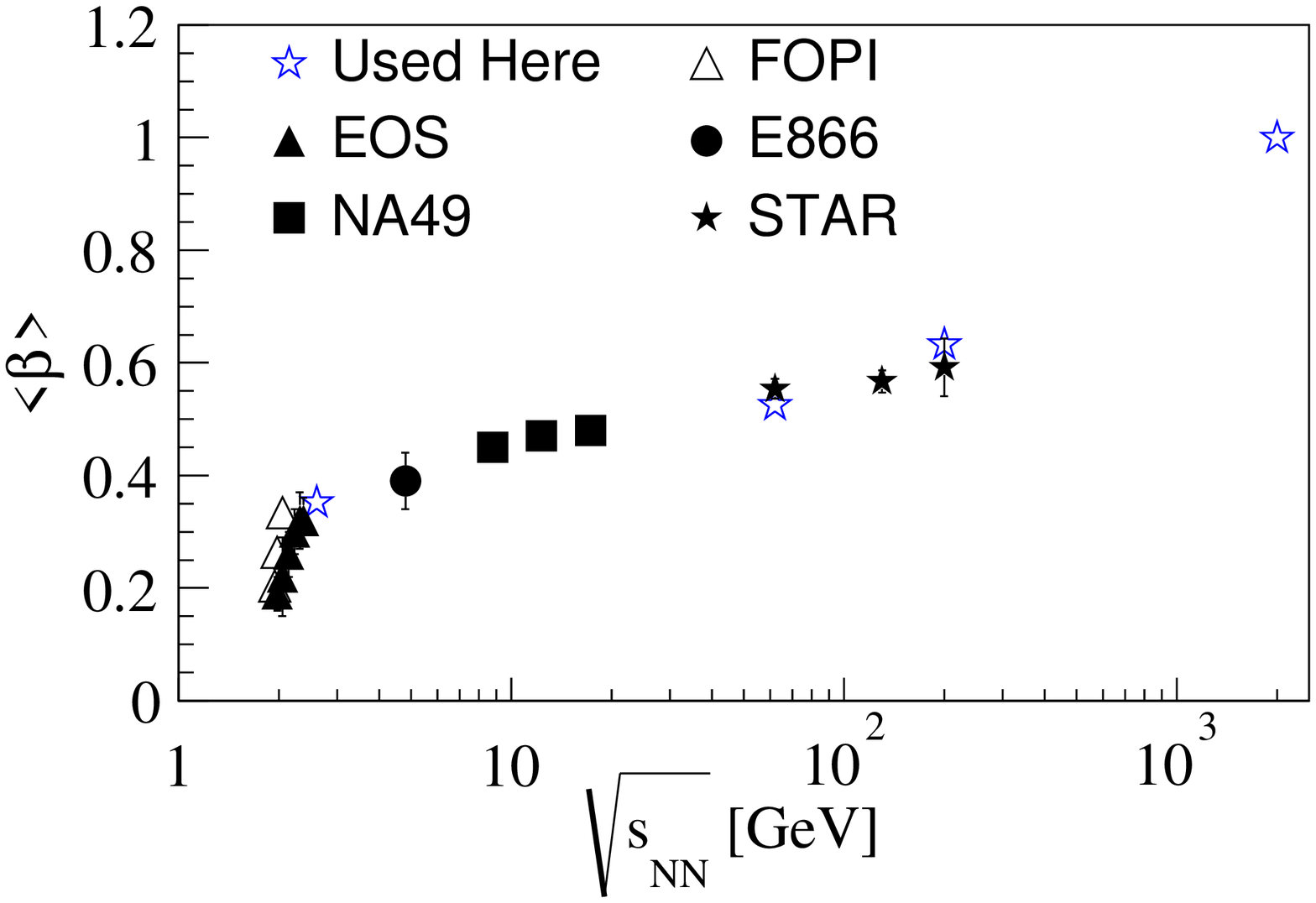}
\caption{\label{fig:beta} (Color online) The radial flow parameters as a function of centre of mass energy $\sqrt{s_{NN}}$;
data are taken from the RHIC-STAR collaboration~\cite{RHIC-STAR-KL200-1,RHIC-STAR-KL200-2}.}
\end{figure}

Transverse momentum ($p_T$) spectrum is a basic observable in
Ultra-relativistic heavy-ion collisions \cite{Wang,Liu}. By fitting the transverse momentum spectra with thermal model, information of the system created in the collisions can be obtained, such as the kinetic freeze-out temperature and the radial flow. If the spectra are identified for different particles species, the ratio between the different particles will give the chemical freeze-out temperature and baryon (strangeness) chemical potentials. In this paper, a blast wave model with chemical equilibrium model is employed to calculate the transverse momentum ($p_T$) spectra of Kaon and $\Lambda$ in nucleus - nucleus collisions at ultra-relativistic energy $\sqrt{s_{NN}}$ from 2.6 GeV to 2.76 TeV corresponding to the FOPI, RHIC and LHC energy range. 

Figure~\ref{fig:pT-FOPI} shows the transverse momentum distribution of $K^{0}$ and $\Lambda$ by the blast wave model compared with the data from FOPI collaboration~\cite{FOPI-KL} in Ni + Ni collisions at  $\sqrt{s_{NN}}\simeq$ 2.6 GeV. Our results can describe the data well. The chemical temperature $T_{ch}$ and the baryon chemical potential $\mu_{B}$ used in this calculation is similar to the ones applied in reference~\cite{dNdy-NPA772-167}. And the kinetic freeze-out parameters of temperature $T_{kin}$ and radial flow $\rho_{0}$ is chosen from experimental results with compilation in the reference~\cite{RHIC-STAR-SYS-KL62}. The production of Kaon and $\Lambda$ here is near the  $\Lambda$ threshold, thus it can be a reference in Lanzhou-Cooling Storage Ring (CSR) facility and High Intensity Accelerator Facility (HIAF) in the near future. Figure~\ref{fig:pT-RHIC62} and~\ref{fig:pT-RHIC200} present the transverse momentum distributions of $K^{+}$ and $\Lambda$ at mid-rapidity in Au + Au central collisions at $\sqrt{s_{NN}}$ = 62.4 GeV and 200 GeV, respectively. The chemical and kinetic freeze-out parameters are extracted from data as in Refs.~\cite{Thermal-PMB-1,Thermal-PMB-2,RHIC-STAR-SYS-KL62}. And for LHC energy the chemical freeze-out parameters and kinetic freeze-out temperatures were already discussed in our previous study~\cite{SONG-LHC-PRC} and figure~\ref{fig:pT-LHC} presents the $p_{T}$ distribution of $K^{+}$ and $\Lambda$ in Pb+Pb central collisions at $\sqrt{s_{NN}}\simeq$ 2.76 TeV. Our calculated results are in good agreement with the $p_{T}$ distribution from experiment. From figure~\ref{fig:pT-FOPI}, \ref{fig:pT-RHIC62}, \ref{fig:pT-RHIC200}, and~\ref{fig:pT-LHC}, it can be found that the transverse momentum distributions of $\Lambda$ is harder than those of Kaon. The slope (effective temperature) of the momentum distribution is related to both the radial flow parameter $\langle \beta \rangle$ and the mass of particles ($m_{0}$), i.e.  $T_{eff}=T_{kin}+\frac{1}{2}m_{0}\langle \beta \rangle^{2}$~\cite{Teff_beta_m0-1,Teff_beta_m0-2,Teff_beta_m0-3}. The radial flow will push the heavier particle to high transverse momentum, therefore the stiffer spectrum emerges for $\Lambda$ in contrast with Kaon.

As discussed above the collective properties of the hot and dense matter created in ultra-relativistic heavy-ion collisions at freeze-out stage can be calculated through equation (\ref{eq-betaT}). Figure~\ref{fig:beta} shows the radial flow parameter $\langle \beta \rangle$ as a function of centre of mass energy $\sqrt{s_{NN}}$. It shows an increasing trend of the radial flow with the increasing of centre of mass energy $\sqrt{s_{NN}}$. Namely the expanding velocity of the collision system in radial direction is larger at higher energy. From figure~\ref{fig:beta} the radial flow used in this calculation is consistent with those extracted from the experimental results~\cite{RHIC-STAR-SYS-KL62}. It implies that these parameters are also valid for strangeness hadrons.

The resonance production rate is high in nucleus-nucleus collisions at relativistic energy. The hadrons decayed from resonances will pollute the $p_T$ distribution of hadrons produced in the collision~\cite{RHIC-STAR-SYS-KL62}. However, it is difficult to measure the effect precisely in experiment due to the wide species of resonance. This can be done in model calculation, though. In our previous work~\cite{SONG-LHC-PRC} this effect for $\pi$, K and p is discussed in details in the LHC energy range. We show the results for strangeness hadron in different energy from figure~\ref{fig:pT-FOPI},~\ref{fig:pT-RHIC62},~\ref{fig:pT-RHIC200} and~\ref{fig:pT-LHC} in this paper. In this calculation, Kaon's and $\Lambda$'s resonances are all included. The kaons (decayed from resonances) are mainly from $K^{*}(892)$, $\phi(1020)$ and other heavy resonances which their masses are heavier than that of 2 Kaons, and $\Lambda$'s (decayed from resonance) are mainly from $\Sigma$'s and $\Xi$'s resonances. From Eq.(~\ref{eq:ni}) and the mass of those particles mentioned above, one can estimate that the production rate from resonance decay of $\Lambda$s will be more than that of Kaons. And Table~\ref{tab:percentDecay} summarises our estimation for this effect. The percent of $K^{+}$($\Lambda$) from resonance decay to total $K^{+}$($\Lambda$) is increased with the increasing of centre of mass energy $\sqrt{s_{NN}}$. The percentage for $K^{+}$ is higher than that for $\Lambda$. And the effect from resonance decay to $K^{+}$ and $\Lambda$ is more significant at RHIC and LHC energy than the ones at low beam energy.

\begin{table*}
\caption{ \label{tab:percentDecay} Percent of $K^{+}$ and $\Lambda$ from resonance decay in total $K^{+}$ and $\Lambda$ vs. centre of mass energy $\sqrt{s_{NN}}$. }
\begin{ruledtabular}
\begin{tabular}{lllll}
$\sqrt{s_{NN}}$ (GeV) & 2.6 & 62 & 200 & 2000\\
\hline
$K^{+}$(decayed)/$K^{+}(total)[\%]$ & 0.2 & 44.4 & 44.6 & 46.9\\
$\Lambda$(decayed)/$\Lambda(total) [\%]$ & 19.5 & 73.8 & 74.1 & 74.9

\end{tabular}
\end{ruledtabular}
\end{table*}

\section{Summary}

The blast wave model with a chemical equilibrium model is employed to calculate the production of Kaon and $\Lambda$ in nucleus-nucleus collisions at relativistic energy. The baryon chemical potential $\mu_{B}$ and chemical freeze-out temperature $T_{ch}$ are consistent with those extracted from experiments~\cite{RHIC-STAR-SYS-KL62} and used in other model calculations~\cite{dNdy-NPA772-167}. The kinetic freeze-out properties are also discussed such as kinetic freeze-out temperature $T_{kin}$
and radial flow parameter $\rho_{0}$. These kinetic freeze-out parameters are in good agreement with the ons from experimental results and are found to be valid for strangeness hadrons (Kaon and $\Lambda$). And the resonance decay effect on strangeness production is also studied in details in this paper. In summary, the beam energy dependence of Kaon and $\Lambda$ production can be a good probe to study the properties of the dense matter created in ultra-relativistic heavy-ion collisions. Our results  shed light on the beam energy scan at RHIC and the academic activity related to strangeness production at Lanzhou-CSR facility.

This work was supported in part by the Major State Basic Research
Development Program in China under Contract No. 2014CB845400, the
National Natural Science Foundation of China under contract Nos. 11421505, 
11035009, 11220101005,  11105207, 11275250, 11322547, U1232206, the Knowledge
Innovation Project of the Chinese Academy of Sciences under Grant
No. KJCX2-EW-N01.


\begin{thebibliography}{35}

\bibitem{QCD-QGP} F. Karsch, "Lattice results on QCD thermodynamics", Nucl. Phys. A \textbf{698}, 199c (2002).

\bibitem{RHICWithePaper-1} I. Arsene \textit{et al.} (BRAHMS Collaboration), "Quark-gluon plasma and color glass condensate at RHIC? The perspective from the BRAHMS experiment", Nucl. Phys. A \textbf{757}, 1
(2005).
\bibitem{RHICWithePaper-2} B. B. Back \textit{et al.} (PHOBOS Collaboration), "The PHOBOS perspective on discoveries at RHIC", 
\textit{ibid}. A \textbf{757}, 28 (2005).
\bibitem{RHICWithePaper-3} J. Adames \textit{et al.}
(STAR Collaboration), "Experimental and theoretical challenges in the search for the quark-gluon plasma: The STAR Collaboration's critical assessment of the evidence from RHIC collisions", \textit{ibid}. A \textbf{757}, 102 (2005).
\bibitem{RHICWithePaper-4} K. Adcox \textit{et al.} (PHENIX Collaboration), "Formation of dense partonic matter in relativistic nucleus-nucleus collisions at RHIC: Experimental evaluation by the PHENIX Collaboration", \textit{ibid}. A
\textbf{757}, 184 (2005).


\bibitem{Tian}J. Tian, J. H. Chen, Y. G. Ma, X. Z. Cai, F. Jin, 
G. L. Ma,  S. Zhang and C. Zhong, "Breaking of the number-of-constituent-quark scaling for identified-particle elliptic flow as a signal of phase change in low-energy data taken at the BNL Relativistic Heavy Ion Collider (RHIC)",  Physical Review C {\bf 79}, 067901 (2009).

\bibitem{LiuFM}F. M. Liu, 	"Explore QCD phase transition with thermal photons",  Nuclear  Science and Techniques {\bf 24}, 050524 (2013).

\bibitem{Ko}C. M. Ko, L. W. Chen, V. Greco, F. Li, Z. W. Lin,
S. Plumari, T. Song and J. Xu, "Mean-field effects on matter and antimatter elliptic flow ",  Nuclear  Science and Techniques  {\bf 24}, 050525 (2013).

\bibitem{Mohanty}B. Mohanty [STAR Collaboration],"STAR experiment results from the beam energy scan program at RHIC", Journal of Physics G {\bf 38}, 124023 (2011).

\bibitem{Phi}B. I. Abelev {\it et al.} (STAR Collaboration) , "Partonic flow and $\phi$-meson production in Au+Au collisions at $\sqrt{s_{NN}}$ = 200 GeV",  Physical Review Letters {\bf 99}, 112301 (2007).

\bibitem{MaYG}Y. G. Ma, "$\phi$-meson production and partonic collectivity at RHIC", Journal of Physics G: Nucl. Part. Phys. {\bf 32}, S373 (2006).

\bibitem{Zhang2010}S. Zhang, J.H. Chen, H. Crawford, D. Keane, Y.G. Ma, Z.B. Xu, "Searching for onset of deconfinement via hypernuclei and baryon-strangeness correlations", Physics Letters B
{\bf 684}, 224 (2010).

\bibitem{ChenJH2008}J. H. Chen, F. Jin, D. Gangadharan, X. Z. Cai, H. Z. Huang, and Y. G. Ma, "Parton distributions at hadronization from bulk dense matter produced in
Au+Au collisions at $\sqrt{s_{NN}}$ = 200 GeV", Physical Review C {\bf 78}, 034907 (2008).

\bibitem{FOPI-KL} M. Merschmeyer  {\it et al.}  (FOPI Collaboration), "$K^{0}$ and $\Lambda$ production in Ni+Ni collisions near threshold", Physical Review C {\bf 76}, 024906 (2007).

\bibitem{KaoS-KL} A. F\"orster  {\it et al.} (KaoS Collaboration), "Production of $K^{+}$ and of $K^{-}$ mesons in heavy-ion collisions from 0.6 A to 2.0 A GeV incident energy", Physical Review C {\bf 75}, 024906 (2007).

\bibitem{RHIC-STAR-SYS-KL62} B. I. Abelev  {\it et al.} (STAR Collaboration), "Systematic measurements of identified particle spectra in pp, d + Au, and Au + Au collisions at the STAR detector", Physical Review C {\bf  79}, 034909 (2009).

\bibitem{RHIC-STAR-KL62} M. M. Aggarwal  {\it et al.} (STAR Collaboration), "Strange and multistrange particle production in Au+Au collisions at $\sqrt{s_{NN}}$ = 62.4 GeV", Physical Review C {\bf 83}, 024901 (2011).

\bibitem{RHIC-STAR-KL200-1} J. Adams  {\it et al.} (STAR Collaboration), "Identified Particle Distributions in pp and Au + Au Collisions at  $\sqrt{s_{NN}}$ = 200 GeV", Physical Review Letters, Vol. 92, 112301 (2004).
\bibitem{RHIC-STAR-KL200-2} G. Agakishiev  {\it et al.} (STAR Collaboration), "Strangeness Enhancement in Cu-Cu and Au-Au Collisions at $\sqrt{s_{NN}}$ = 200 GeV", Physical Review Letters {\bf 108}, 072301 (2012).

\bibitem{LHC-ALICE-KL-1} B. M\"uller, J. Schukraft, B. Wys\l ouch, "First Results from Pb+Pb Collisions at the LHC", Annual Review of Nuclear and Particle Science {\bf 62}, 361 (2012).
\bibitem{LHC-ALICE-KL-2} D.D. Chinellato for the ALICE Collaboration, "Strange and Multi-Strange Particle Production in ALICE", Journal of Physics: Conference Series {\bf 446}, 012055 (2013).

\bibitem{LHC2-1}K. Aamodt, N. Abel, U. Abeysekara {\it et al.}, "Charged-particle multiplicity measurement in proton–proton collisions at \(\sqrt{s}=7\)  TeV with ALICE at LHC", European Physics Journal C {\bf 68}, 345 (2010).
\bibitem{LHC2-2}H, M. Wang, Z. Y. Hou, X. J. Sun,  "Hadron multiplicities in p+p and p+Pb collisions at the LHC",  Nuclear  Science and Techniques  {\bf 25}, 040502 (2014).

\bibitem{SONG-LHC-PRC} S. Zhang, L. X. Han, Y. G. Ma  {\it et al.}, "Production and ratio of $\pi$, K, p, and $\Lambda$ in Pb + Pb collisions at $\sqrt{s_{NN}}$ = 2.76 TeV", Physical Review C {\bf 89}, 034918 (2014).

\bibitem{BLWave} E. Schnedermann, J. Sollfrank,  U. Heinz, "Thermal phenomenology of hadrons from 200A GeV S+S collisions", Physical Review C \textbf{48}, 2462 (1993).

\bibitem{ALICE-chPCDepen} B. Abelev \textit{et al.} (ALICE Collaboration), "Centrality dependence of $\pi$, K, and p production in Pb-Pb collisions at $\sqrt{s_{NN}}$=2.76 TeV ", Physical Review C \textbf{88}, 044910 (2013).

\bibitem{BLWave-Fabrice} F. Reti\`ere, M. A. Lisa, "Observable implications of geometrical and dynamical aspects of freeze-out in heavy ion collisions", Physical Review C \textbf{70}, 044907 (2004).

\bibitem{DRAGON} B. Tom\'a\v sik, "DRAGON: Monte Carlo generator of particle production from a fragmented fireball in ultrarelativistic nuclear collisions", Comput. Phys. Commun. \textbf{180}, 1642 (2009).

\bibitem{THERMINATOR-1} M. Chojnacki, A. Kisiel, W. Florkowski, W. Broniowski, "THERMINATOR 2: THERMal heavy IoN generATOR 2", Comput. Phys. Commun. \textbf{183}, 746 (2012).
\bibitem{THERMINATOR-2} A. Kisiel, T. Ta\l u\'c, W. Broniowski,  W. Florkowski, "THERMINATOR: THERMal heavy-IoN generATOR", Comput. Phys. Commun. \textbf{174}, 669 (2006).

\bibitem{Thermal-EQ-1} P. Braun-Munzinger, J. Stachel, J. P. Wessels, N. Xu, "Thermal equilibration and expansion in nucleus-nucleus collisions at the AGS", Phys. Lett. B \textbf{344}, 43 (1995).
\bibitem{Thermal-EQ-2} P. Braun-Munzinger, J. Stachel, J. P. Wessels,  N. Xu, "Thermal and hadrochemical equilibration in nucleus-nucleus collisions at the SPS", Phys. Lett. B \textbf{365}, 1 (1996).
\bibitem{Thermal-EQ-3} P. Braun-Munzinger, I. Heppe,  J. Stachel, "Chemical equilibration in Pb+Pb collisions at the SPS", Physics Letters B \textbf{465}, 15 (1999).
\bibitem{Thermal-EQ-4}  N. Xu, M. Kaneta, "Hadron freeze-out conditions in high energy nuclear collisions", Nuclear Physics A \textbf{698}, 306 (2002).

\bibitem{Thermal-PMB-1} P. Braun-Munzinger, K. Redlich,  J. Stachel, "Particle Production in Heavy Ion Collisions", arXiv:nucl-th/0304013v1.
\bibitem{Thermal-PMB-2} A. Andronic, P. Braun-Munzinger and J. Stachel, "Thermal hadron production in relativistic nuclear collisions: The hadron mass spectrum, the horn, and the QCD phase transition", Physics Letters B \textbf{673}, 142 (2009).

\bibitem{dNdy-NPA772-167} A. Andronic, P. Braun-Munzinger, J. Stachel, "Hadron production in central nucleus-nucleus collisions at chemical freeze-out", Nuclear Physics A {\bf 772}, 167 (2006).

\bibitem{LNi-CPC-35-741} S. Zhang, J. H. Chen, Y. G. Ma  {\it et al.}, "Hypertriton and light nuclei production at $\Lambda$-production subthreshold energy in heavy-ion collisions", Chinese Physics C {\bf  35}, 741 (2011).

\bibitem{Wang}X.-N. Wang, M. Gyulassy, "Hijing: A Monte Carlo model for multiple jet production in pp, pA, and AA collisions", Physical Review D {\bf 44}, 3501 (1991).

\bibitem{Liu} Fu-Hu Liu, Ya-Qin Gao, Tian Tian, and Bao-Chun Li, "Transverse Momentum and Pseudorapidity Distributions of Charged Particles and Spatial Shapes of Interacting Events in Pb-Pb Collisions at 2.76 TeV",
Advance in High Energy Physics {\bf 2014}, 725739 (2014).

\bibitem{Teff_beta_m0-1} R. Scheibl and U. Heinz, "Coalescence and flow in ultrarelativistic heavy ion collisions", Phys. Rev. C \textbf{59}, 1585 (1999).
\bibitem{Teff_beta_m0-2} U. Heinz, "The Little Bang: Searching for quark-gluon matter in relativistic heavy-ion collisions", Nuclear Physics A \textbf{685}, 414c (2001).
\bibitem{Teff_beta_m0-3} N. Xu and M. Kaneta, "Hadon freeze-out conditions in high energy nuclear collisions", Nuclear Physics A \textbf{698}, 306c (2002).

\end{thebibliography}
\end{document}